# Wavelength-by-wavelength temperature-independent thermal radiation utilizing an insulator–metal transition

Jonathan King[1*], Alireza Shahsafi[1*], Zhen Zhang[3], Chenghao Wan[1,2], Yuzhe Xiao[1], Chengzi Huang[3], Yifei Sun[3], Patrick J. Roney[1], Shriram Ramanathan[3] and Mikhail A. Kats[1,2,4]

[1]Department of Electrical and Computer Engineering, University of Wisconsin-Madison

[2]Department of Materials Science and Engineering, University of Wisconsin-Madison

[3]School of Materials Engineering, Purdue University, West Lafayette, Indiana

[4]Department of Physics, University of Wisconsin-Madison

[*]Authors contributed equally

## Abstract

Both the magnitude and spectrum of the blackbody-radiation distribution change with temperature. Here, we designed the temperature-dependent spectral emissivity of a coating to counteract all the changes in the blackbody-radiation distribution over a certain temperature range, enabled by the nonhysteretic insulator-to-metal phase transition of $SmNiO_3$. At each wavelength within the long-wave infrared atmospheric-transparency window, the thermal radiance of our coating remains nearly constant over a temperature range of at least 20 °C. Our approach can conceal thermal gradients and transient temperature changes from infrared imaging systems, including those that discriminate by wavelength, such as multispectral and hyperspectral cameras.

The imaging of thermal radiation using infrared cameras forms the basis of night vision, and enables the measurement of temperature profiles on the surfaces of remote objects. As such, the obfuscation of temperature information provided by infrared imaging is an ongoing research area, with approaches being explored that manipulate one or more of surface temperature, emissivity, and scattering. Active control of temperature [1] or emissivity [2]–[5] can dynamically alter thermal signatures using external power and control circuitry, which in the best case can render an object invisible to an infrared camera by displaying an infrared image of what is behind the object. Passive systems that do not require external power typically cannot dynamically reproduce infrared images that are behind an object, but they can provide a variety of other capabilities. These passive systems include scattering and absorbing screens that block radiation from an object and equilibrate with the environment, and thus can mimic ambient radiation [6]; enclosures with position-dependent effective thermal conductance that enables the replication of a background temperature distribution onto the enclosure exterior [7]; and coatings with non-trivial temperature-dependent emissivities that can make the emitted thermal radiation of an object not depend on temperature [8]. This last class of infrared-concealment systems—passive coatings based on temperature-dependent emissivity—is the focus of this paper.

Such coatings require the incorporation of materials whose infrared properties change strongly with relatively small changes in temperature. For this purpose, phase-transition materials such as vanadium dioxide ($VO_2$) [9], germanium-antimony-tellurium (GST) [10][11], and samarium nickel oxide ($SmNiO_3$) [8] have been explored. Phase-transition materials have been leveraged to enable dynamic thermal emission for a variety of applications including smart windows [12], satellite thermal regulation [13]–[15], and passive daytime cooling [16], in addition to thermal camouflage [2][3].

Among the materials that have been explored for thermal-emission engineering, $SmNiO_3$ is a newcomer, but its unique combination of characteristics provides functionality not found in other material systems. The optical properties of $SmNiO_3$ films can exhibit a strong and steady evolution over tens of degrees (from 25 °C to 140 °C [8]), with especially large changes in the mid- and long-wave infrared, i.e., the wavelength ranges commonly used by infrared imaging systems. Unlike other materials with insulator-to-metal transitions (IMTs), thin films of $SmNiO_3$ can be synthesized such that the hysteresis often associated with IMTs appears to be negligible or absent [17][18].

In 2019, our groups demonstrated the use of $SmNiO_3$ films to realize "zero-differential thermal emitters", or ZDTEs, in which the thermal emission from a coating has a roughly constant power with respect to temperature, when integrated over the wavelength range of the long-wave infrared atmospheric transparency window (8–14 μm) [8]. ZDTEs can be used to conceal information typically present in heat signatures picked up by off-the-shelf infrared cameras. However, in these ZDTEs, only the wavelength-integrated radiated power remains constant with temperature, and the films cease to appear zero-differential when imaged using a hyperspectral camera or even bandpass filters positioned in front of a conventional infrared camera[1].

Here, we designed and demonstrated "zero-differential spectral emitters" (ZDSEs), which are passive emission coatings that feature temperature-independent spectral radiance, and can thus conceal information even from hyperspectral infrared cameras that discriminate by wavelength. In other words, we designed the temperature-dependent emissivity of a coating to counteract the intrinsic temperature-dependent changes at each wavelength in the blackbody thermal-emission spectrum.

---

[1] We note that existing active infrared camouflage systems that are designed to display infrared images of what is behind an object have a similar limitation to our ZDTEs in that they modulate the emitted power rather than the full emitted spectrum.

**Design and optimization**

Any object emits thermal radiation, with the emitted temperature-dependent spectral radiance given by [19]

$$I(\lambda, T) = \varepsilon(\lambda, T) I_{BB}(\lambda, T) \qquad (1),$$

where $\varepsilon(\lambda, T)$ is the object's spectral emissivity and $I_{BB}(\lambda, T)$ is the spectral radiance of a blackbody, which is given by Planck's law [20] [Fig. 1(a)]:

$$I_{BB}(\lambda, T) = \frac{2hc^2}{\lambda^5} \frac{1}{e^{\frac{hc}{\lambda k_B T}} - 1}. \qquad (2)$$

Here, $h$ is Planck's constant, $k_B$ is Boltzmann's constant, $c$ is the free-space speed of light, $\lambda$ is free-space wavelength, and $T$ is the absolute temperature.

To achieve temperature-independent spectral radiance, we require $dI(\lambda, T)/dT = 0$ for every wavelength of interest. The solution to this differential equation requires the emissivity to be in the form (*SI Appendix*, section 1)

$$\varepsilon(\lambda, T) = f(\lambda)\left(e^{\frac{hc}{\lambda k_B T}} - 1\right), \qquad (3)$$

where $f(\lambda)$ is any function of $\lambda$. Eqn. (3) provides a substantial degree of freedom when designing coatings with temperature-independent spectral radiance, since $f(\lambda)$ is an arbitrary function [Fig. 1(b)].

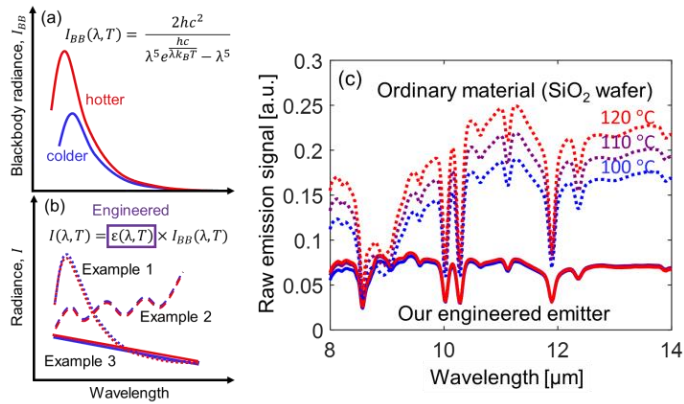

**Figure 1) (a)** Blackbody spectral radiance, $I_{BB}(\lambda, T)$, at two arbitrary temperatures. **(b)** Several examples of temperature-dependent spectral radiance where $dI(\lambda, T)/dT = 0$ is achieved by engineering the emissivity. (c) The main experimental result of this paper: an emitter engineered to achieve $dI(\lambda, T)/dT = 0$ across the 8 – 14 μm from $T$

= 100 °C to $T$ = 120 °C compared to a reference SiO$_2$ wafer (the sharp features are due to absorption by ambient gases present in the lab on the day of the measurement).

Our goal was therefore to implement a form of $\varepsilon(\lambda, T)$ that satisfied Eqn. (3) over the long-wave atmospheric transparency range (8 to 14 µm) using SmNiO$_3$. We hypothesized that the design process would be simpler if the substrate for the SmNiO$_3$ film was reflective and did not have any abrupt features in its spectrum in the 8 – 14 µm range. Indium tin oxide (ITO) is well-suited for this purpose, which depending on its stoichiometry can be reflective and featureless in our wavelength range of interest [21].

Based on this argument, we selected our structure to be a thin-film assembly of SmNiO$_3$, ITO, and a substrate [Fig. 2(a)], with the thicknesses of the films ($d_{SNO}$ for SmNiO$_3$ and $d_{ITO}$ for ITO) to be determined by an optimization process. We chose the substrate to be soda-lime glass due to the commercial availability of inexpensive ITO films on these substrates.

To find the ideal thicknesses of the layers of the SmNiO$_3$ and ITO films, we used the transfer-matrix method to calculate the reflectance for different layer thicknesses, which we converted to emissivity using Kirchhoff's law, and then combined with Planck's law to determine the emitted spectral radiance (Fig. 2). We selected a figure of merit (to be minimized) that reasonably captures how close an emitter is to a ZDSE around a given temperature $T$:

$$\Phi(T) = \int_{\lambda_1}^{\lambda_2} 1/I(\lambda, T) \cdot |\partial I(\lambda, T)/\partial T| \, d\lambda \quad (4)$$

Note that a simpler figure of merit that omits the $1/I(\lambda, T)$ term in Eqn. (4) can also be used to find solutions with small absolute changes of radiance with temperature, but these solutions would often simply be "trivial" low-emissivity designs; note that an object with $\varepsilon(\lambda, T) = 0$ is always zero-differential. Instead, the full Eqn. (4) can be used to find nontrivial ZDSE designs with emissivity values much greater than 0.

In these calculations, we used the refractive indices of SmNiO$_3$ that we measured previously [8] [Fig. 2(b) and Fig. 2(c)]. The refractive indices of ITO were derived from Drude-theory calculations, assuming a free-carrier density of $n_e = 8 \times 10^{20}$ cm$^{-3}$ (see calculations with other $n_e$ values in *SI*, section 3). Because soda-lime glass is mostly SiO$_2$ and we expected most light not to reach the substrate anyway, we used the complex refractive index of SiO$_2$ glass [22] to describe the soda-lime substrate in the calculations. We chose to perform our ZDSE optimization around 110 °C based on the location of the phase transition and availability of refractive-index data.

The resulting map of the figure of merit $\Phi(T = 110 \,°C)$ is shown in Fig. 2(d). The device achieves good performance (i.e., minimal $\Phi(T = 110 \,°C)$) at $d_{SNO} > 200$ nm and almost any value of $d_{ITO}$, though $d_{ITO}$ can be used as a fine-tuning knob to slightly change $\Phi$. The optimal parameters were $d_{SNO} \sim 320$ nm and $d_{ITO} \sim 500$ nm. Note that the minima in Fig. 2(d) are broad, so for ease of fabrication, we chose a slightly suboptimal set of parameters, with a thinner film of SmNiO$_3$, to make it easier to deposit [Fig. 2(d)].

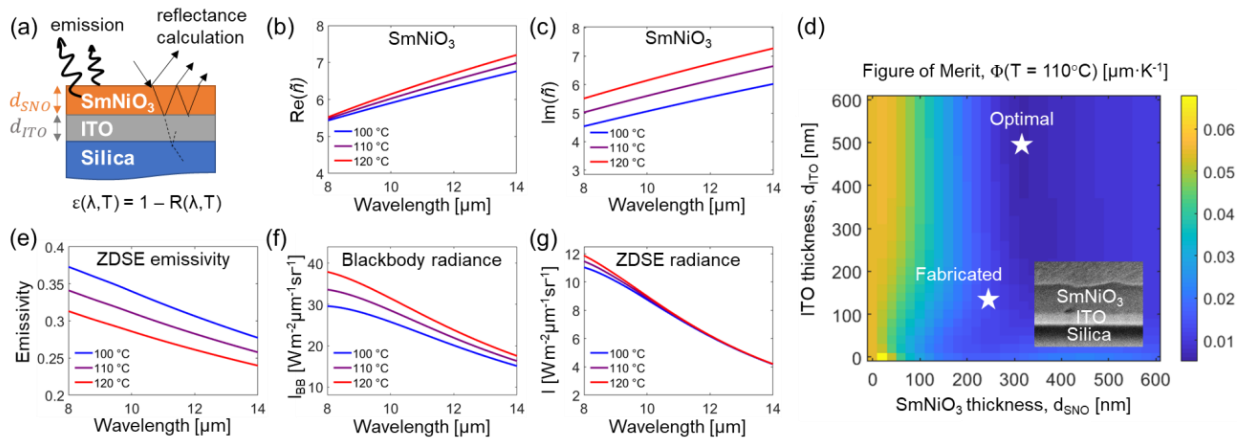

**Figure 2) (a)** Thin-film assembly consisting of SmNiO$_3$ and ITO films on top of a soda-lime glass substrate (assumed to be pure silica in the simulations). Spectral emissivity is calculated as $1 - R(\lambda, T)$ using the transfer-matrix method. **(b)** Real and **(c)** imaginary parts of the complex refractive index of the SmNiO$_3$ film from $T$ = 100 °C to $T$ = 120 °C [8], which served as input for our transfer-matrix calculations. The indices at $T$ = 110 °C were interpolated from data at $T$ = 100 °C and $T$ = 120 °C. **(d)** Calculated optimization space based on the integrated normalized absolute rate change of spectral radiance with respect to temperature [figure of merit, $\Phi(T)$]. The optimal structure dimensions and actual dimensions of our fabricated device are marked by white stars. The inset is an SEM cross-section of the fabricated structure. **(e)** Spectral emissivity of the optimal zero-differential spectral emitter (ZDSE) from $T$ = 100 °C to $T$ = 120 °C for 8 μm < $\lambda$ < 14 μm. **(f)** Spectral radiance of a blackbody from $T$ = 100 °C to $T$ = 120 °C. **(g)** Spectral radiance of the optimal ZDSE with SmNiO$_3$ and ITO thicknesses of 320 and 500 nm, respectively. Radiance is calculated by multiplying the spectral emissivity by blackbody spectral radiance for each temperature.

**Fabrication and experiments**

We fabricated a sample having dimensions that were within the high-performance design space [dark blue region in Fig. 2(g)]. The thicknesses of SmNiO$_3$ and ITO were $d_{SNO} = 234 \pm 18$ nm and $d_{ITO} = 120 \pm 5$ nm, respectively, measured after fabrication using scanning electron microscopy (SEM) cross section images (details and images provided in *SI Appendix*, section 4). Our ITO-coated soda-lime glass substrate was purchased from Sigma Aldrich, and had a resistivity of 8-12 Ω/sq. SmNiO$_3$ thin films were magnetron co-sputtered from Ni and Sm targets at room temperature. DC power of 90 W was applied to Ni target, and RF power of 170 W was applied to Sm target. The stoichiometric ratio of Sm and Ni was calibrated with energy dispersive spectroscopy. During deposition, the pressure of the chamber was kept at 5 mTorr with flow of Ar (40 sccm) and O$_2$ (10 sccm) gases. To form the perovskite phase, the as-grown films were annealed at 500 °C in pure oxygen gas at 1400 psi for 24 hours.

The reflectance of our fabricated ZDSE was measured using Fourier-transform spectroscopy (FTS, Hyperion 2000 microscope coupled to Bruker Vertex 70 spectrometer) and a temperature-controlled heat stage, as shown in Fig. 3(a), and the emissivity was obtained via Kirchhoff's law for opaque, non-scattering samples (*i.e.*, $\varepsilon = 1 - R$) [8]. The emissivity of the fabricated ZDSE [Fig. 3(a)] roughly resembles our optimized theoretical emissivity shown in Fig. 2(e), thus approaching the intended spectral and thermal dependencies that satisfy the criterion of $dI(\lambda, T)/dT = 0$. As a result, the figure of merit [Eqn. (4)] of our fabricated ZDSE is $\Phi(T) \sim 0.007$ µm·K$^{-1}$, which is close to the theoretical minimum of $\Phi(T) \sim 0.006$ µm·K$^{-1}$. The nearly temperature-independent spectral radiance was observed directly by measuring the raw emission signal of the heated sample using our FTS setup described previously [23], as shown in Fig. 3(b). The same sample was also measured on another day, and directly compared to a reference SiO$_2$ wafer [Fig. 1(c)], demonstrating its broad temperature-independence at each wavelength within the long-wave infrared.

We also imaged the samples using a long-wave infrared camera (FLIR A325sc) with and without spectral filters between the camera and the sample, to demonstrate temperature independence in various spectral bands. When performing infrared imaging with this type of camera, typically the user inputs a wavelength-averaged emissivity $\varepsilon_{set}$, as well as measurement conditions such as the background temperature, and the camera software then reports a temperature, which we refer to as the "apparent temperature". Because in our sample the actual emissivity changes with temperature, there is ambiguity about what $\varepsilon_{set}$ should be. For Fig. 3(c), we first set our temperature stage to 90 °C, and varied $\varepsilon_{set}$ until the apparent temperature of the sample was roughly 90 °C. We then maintained $\varepsilon_{set}$ at the same value for measurements at elevated temperatures. These values of $\varepsilon_{set}$ are reported in Fig. 3(c). Additional IR images and analysis are provided in *SI Appendix*, section 7.

Additional temperature-dependent reflection measurements with heating and cooling cycles were performed to confirm the non-hysteretic IMT in our SmNiO$_3$ film (*SI Appendix*, section 6). We note that the fabricated structure was not uniform, which is apparent from the infrared images in Fig. 3(c) as well as in additional FTS measurements that we provide in *SI Appendix*, section 5. In the infrared images, specks appear on the surface and become more distinct as temperature increases. The specks are likely defects in the SmNiO$_3$ film or dust particulates on the surface.

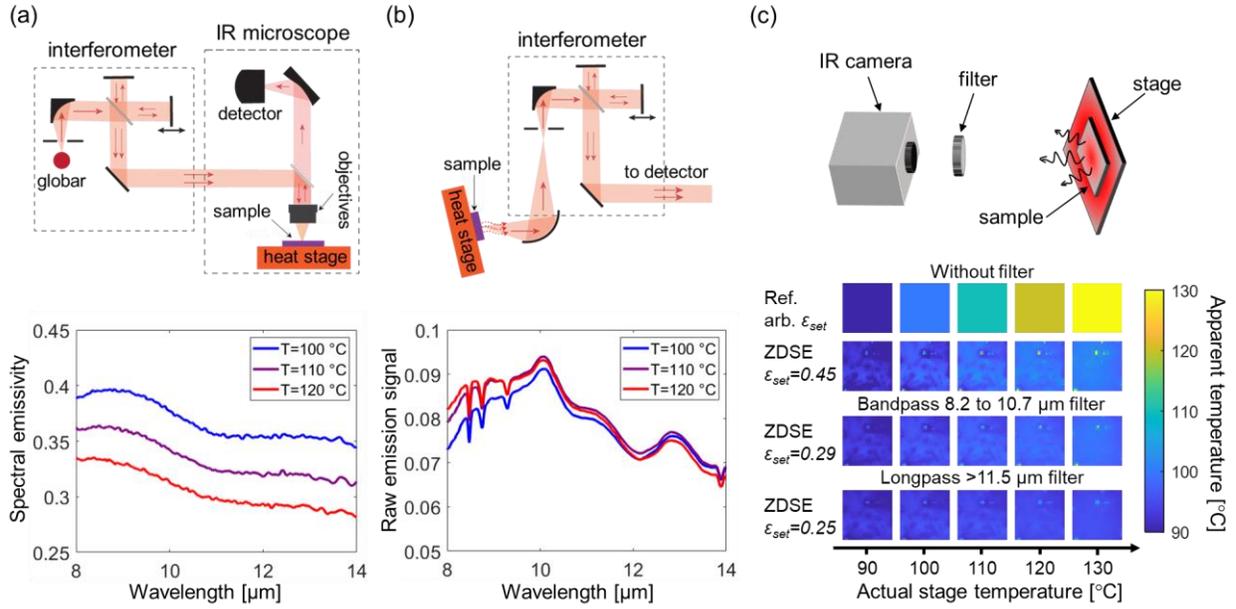

**Figure 3) (a)** (Top) Schematic of reflectance measurements used to obtain spectral emissivity values of our ZDSE and (bottom) resultant spectral emissivity from 100 to 120 °C. **(b)** (Top) Schematic of direct emission measurements and (bottom) the raw spectral emission signal of the ZDSE measured from 100 to 120 °C. The sharp features at 8 < λ < 10 µm are atmospheric absorption lines. **(c)** (Top) Infrared imaging setup and (bottom) infrared images of the ZDSE with and without spectral filters. The "apparent temperature" is the value reported by the FLIR infrared camera when the input emissivity $\varepsilon_{set}$ was selected such that the apparent temperature matched the stage temperature at 90 °C (see main text for more details).

**Discussion and conclusion**

It is not trivial that a coating can be engineered to precisely cancel the changes in the Planck distribution for all individual wavelengths within a broad wavelength band (a "zero-differential spectral emitter", or ZDSE), over a broad range of temperatures. The emissivity spectra required for a ZDSE must satisfy Kramers-Kronig relations, while using combinations of materials with appropriate temperature-dependent optical properties and micro/nanostructure. However, since there are many solutions to Eqn. (3), we were able to find a practical solution based on the phase transition of samarium nickel oxide ($SmNiO_3$) for the long-wave infrared atmospheric transparency window.

We experimentally demonstrated a ZDSE based on films of $SmNiO_3$ and indium tin oxide (ITO), which achieves temperature-independent spectral radiance from ~90 to ~130 °C in the long-wave atmospheric transparency window of 8 to 14 μm. Future ZDSE designs may employ quaternary alloys such as $Sm_XNd_{1-X}NiO_3$, which can shift the phase transition to lower temperatures compared to $SmNiO_3$ [17], or vanadium dioxide ($VO_2$) where the intrinsic hysteresis of its phase transition can be minimized by doping [24][25].

The ZDSE conceals temperature gradients and transient temperature swings on a surface from infrared detection. More broadly, consideration of not only thermally emitted power but also the spectrum is required for any technology—passive or active—that aims to conceal infrared signatures from multispectral and hyperspectral imaging systems.

**Data Availability**

Data for all the figures is available upon reasonable request.

**Acknowledgements**

MAK acknowledges funding support from ONR N00014-20-1-2297 and NSF 1750341. SR acknowledges funding support from AFOSR grant FA9550-19-1-0351.

# Wavelength-by-wavelength temperature-independent thermal radiation utilizing an insulator–metal transition

Jonathan King[1*], Alireza Shahsafi[1*], Zhen Zhang[3], Chenghao Wan[1,2], Yuzhe Xiao[1], Chengzi Huang[3], Yifei Sun[3], Patrick J. Roney[1], Shriram Ramanathan[3] and Mikhail A. Kats[1,2,4]

[1]Department of Electrical and Computer Engineering, University of Wisconsin-Madison

[2]Department of Materials Science and Engineering, University of Wisconsin-Madison

[3]School of Materials Engineering, Purdue University, West Lafayette, Indiana

[4]Department of Physics, University of Wisconsin-Madison

[*]Authors contributed equally

## 1. General solution for zero-differential spectral emission (ZDSE)

For ease of reading this section, we first repeat relevant elements of the introduction of the main text. All objects radiate, with the emitted temperature-dependent spectral radiance $I(\lambda, T)$ described by blackbody spectral radiance multiplied by a wavelength-dependent emissivity $\varepsilon(\lambda, T)$ [S1]:

$$I(\lambda, T) = \varepsilon(\lambda, T) I_{BB}(\lambda, T) \tag{S1}$$

where blackbody spectral radiance, $I_{BB}(\lambda, T)$, is given by Planck's law [S2]:

$$I_{BB}(\lambda, T) = \frac{2hc^2}{\lambda^5} \frac{1}{e^{\frac{hc}{\lambda k_B T}} - 1} \tag{S2}$$

Here, $h$ is Planck's constant, $k_B$ is Boltzmann's constant, $c$ is the free space speed of light, $\lambda$ is free space wavelength, and $T$ is absolute temperature. Zero-differential spectral emission occurs when at each wavelength:

$$\frac{dI(\lambda, T)}{dT} = 0 \tag{S3}$$

Plugging (S1) into (S3) gives

$$\frac{d(\varepsilon(\lambda, T) I_{BB}(\lambda, T))}{dT} = 0 \tag{S4}$$

Applying the product rule gives

$$\frac{d\varepsilon(\lambda, T)}{dT} I_{BB}(\lambda, T) + \varepsilon(\lambda, T) \frac{dI_{BB}(\lambda, T)}{dT} = 0, \tag{S5}$$

which is a differential equation whose solution is:

$$\varepsilon(\lambda, T) = f(\lambda) \left( \exp\left(\frac{hc}{\lambda k_b T}\right) - 1 \right), \tag{S6}$$

where $f(\lambda)$ is an arbitrary function of $\lambda$.

## 2. Calculation of the figure of merit, $\Phi$

Since the refractive index of SmNiO$_3$ was characterized using ellipsometry measurements taken at discrete temperatures, our figures of merit calculations used finite-difference approximations of derivative terms. For example, we minimized $\Phi(T = 110\ °C)$ according to Eqn. (S7)

$$\Phi(T = 110\ °C) \approx \int_{8\ \mu m}^{14\ \mu m} \frac{1}{I(\lambda, T=110\ °C)} \left| \frac{I(\lambda, T=120\ °C) - I(\lambda, T=100\ °C)}{120\ °C - 100\ °C} \right| d\lambda \tag{S7}$$

Where $I(\lambda, T = 100\ °C)$ and $I(\lambda, T = 120\ °C)$ were calculated according to Eqn. (S1) where $\varepsilon(\lambda, T)$ is calculated according to Kirchhoff's law ($\varepsilon(\lambda, T) = 1 - R(\lambda, T)$) with reflectance calculated via the transfer matrix method. The refractive indices used in the transfer matrix method are from literature values of SmNiO$_3$ [S3], ITO [S4], and silica [S5]. $I(\lambda, T = 110\ °C)$ is calculated similarly, but the refractive indices of SmNiO$_3$ are an interpolation of SmNiO$_3$ indices at 100 and 120 °C (i.e., we assumed that the real part of the refractive index at 110 °C is the average of the real part at 100 and 120 °C, and the imaginary part of the refractive index at 110 °C is the average of the imaginary part at 100 and 120 °C).

## 3. Varying ITO free carrier density

The ITO free-carrier density, and ITO refractive indices by extension, can vary substantially depending on fabrication process [S4]. So, we also repeated our $\Phi(T)$ calculations for a wide range of possible ITO free carrier densities (Fig. S1). Despite the substantial differences in ITO free-carrier density during simulation, an effective structure could be achieved at d$_1$ ~ 300 nm and d$_2$ ~ 300 nm across free carrier densities.

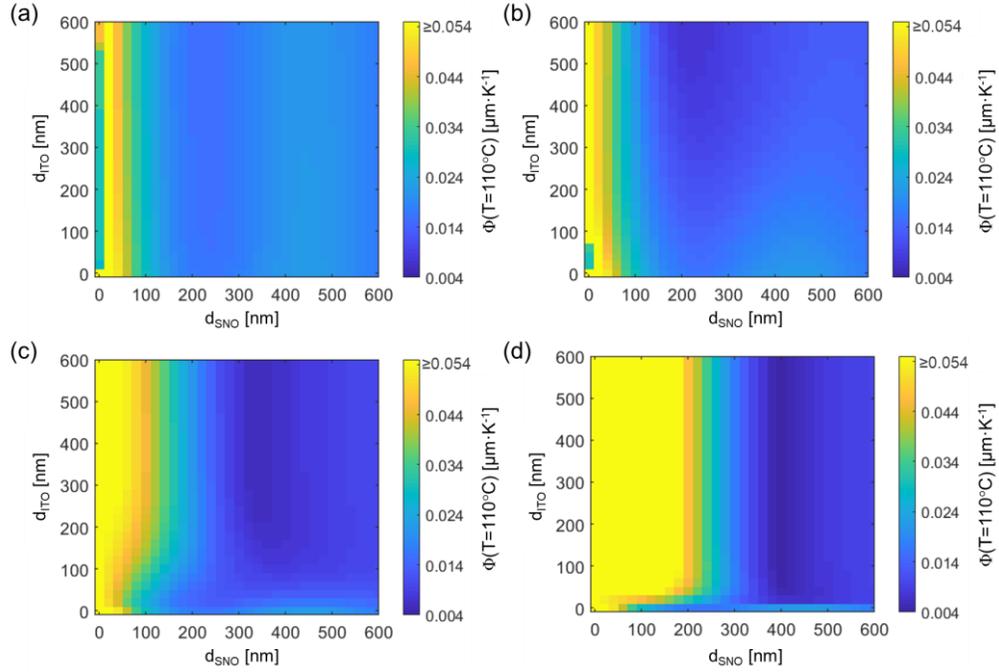

**Figure S1) Integrated normalized absolute rate change of spectral radiance with respect to temperature, $\Phi(T)$.** $\Phi(T)$ assuming an ITO free carrier density of **(a)** $n_e$ = 1E19 cm$^{-3}$, **(b)** $n_e$ = 1E20 cm$^{-3}$, **(a)** $n_e$ = 1E21 cm$^{-3}$ **(a)** $n_e$ = 1E22 cm$^{-3}$

## 4. Cross-sectional SEM imaging

We measured the thicknesses of SmNiO$_3$ and ITO layers by imaging the cross section of the fabricated sample using focused-ion-beam (FIB)-assisted scanning electron microscopy (FIB-SEM, Zeiss Auriga). The cross section was created by a FIB-assisted milling process, as shown in Fig S2(a). To obtain clear images of the SmNiO$_3$ boundaries, we deposited a ~1-µm thick, 10 µm by 2 µm rectangular patch of platinum (Pt) on top of the SmNiO$_3$ and milled through the layers within this patch area, resulting in a clean cross section for SEM imaging [Fig. S2(b)]. The tilted angle of 36° between the SEM beam and the cross section is compensated in the zoomed-in image [Fig. S2(c)]. We measured the thicknesses to be (234 ± 18) nm and (120 ± 5) nm for SmNiO$_3$ and ITO, respectively.

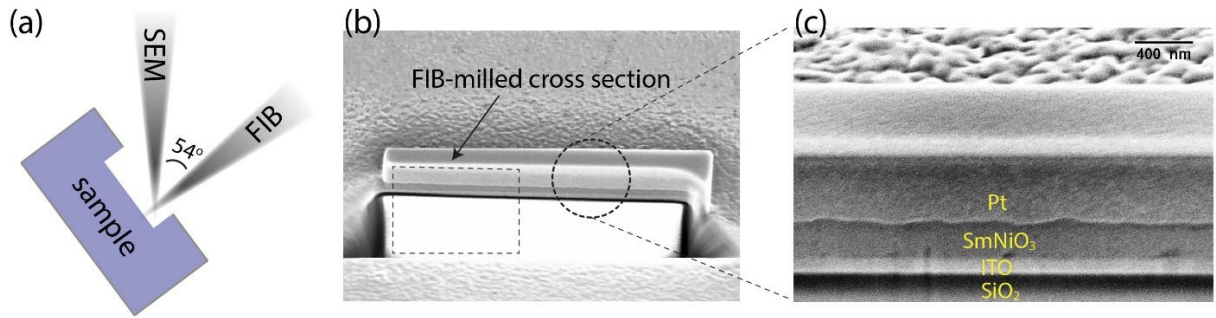

**Figure S2) (a)** A schematic of the SEM imaging on a FIB-milled cross section. **(b)** SEM image of the FIB-milled region. **(c)** Zoomed-in SEM image of the cross section.

## 5. Spatial variability of the emission from sample

We measured the raw emission signal from four different arbitrarily chosen spots on the device at $T = 100$ °C, $T = 110$ °C, $T = 120$ °C, and $T = 130$ °C (the second measured spot did not include the $T = 130$ °C measurement) using the setup featured in the main text. Results are provided in figure in Fig. S3 with the data in Figure S3(c) used in Fig. 3(b) of the manuscript.

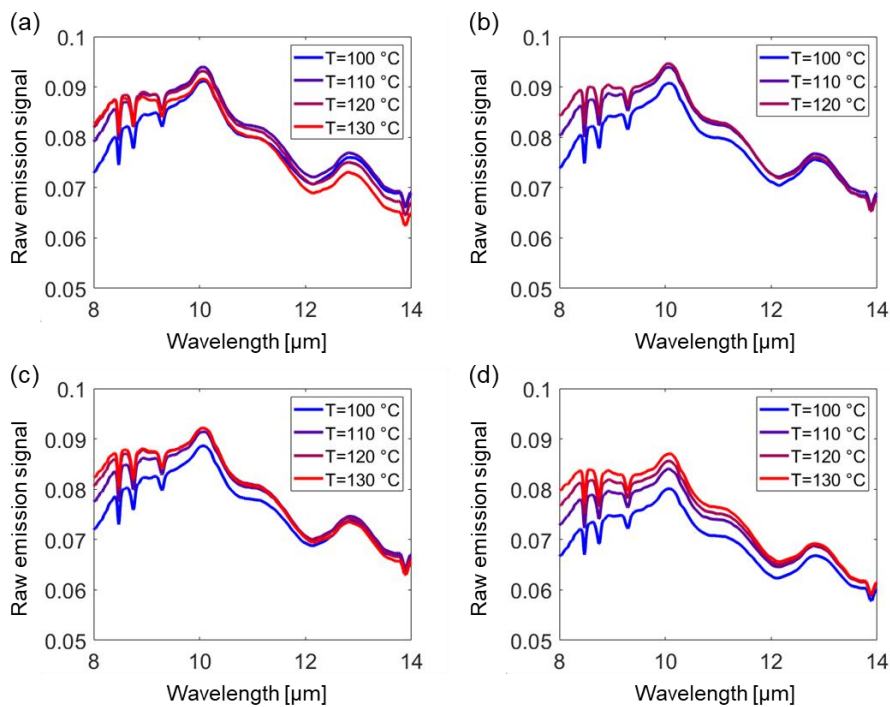

**Figure S3) (a)-(d) Raw emission signal measured at four different arbitrary locations.** Measurement in **(b)** did not include a temperature of T = 130 °C.

## 6. Measurements of hysteresis

In addition to spatial variation, we measured the sample emissivity (via Kirchhoff's law, reflectance measurement with N.A. = 0.4) while temperature cycling to verify a hysteresis-free IMT following the setup shown in Fig. 3(a) of the main text. Through the initial two thermal cycles from 40 °C to 160 °C and back down to 40 °C, the sample appeared to exhibit some hysteresis. However, by the third temperature cycle, the sample exhibited no discernable hysteresis. Our second thermal cycle from 40 °C to 160 °C is shown in Fig. S4. Our third thermal cycle from 40 °C to 160 °C is shown in Fig. S5. Subsequent cycling experiments showed no discernable hysteresis after two cycles.

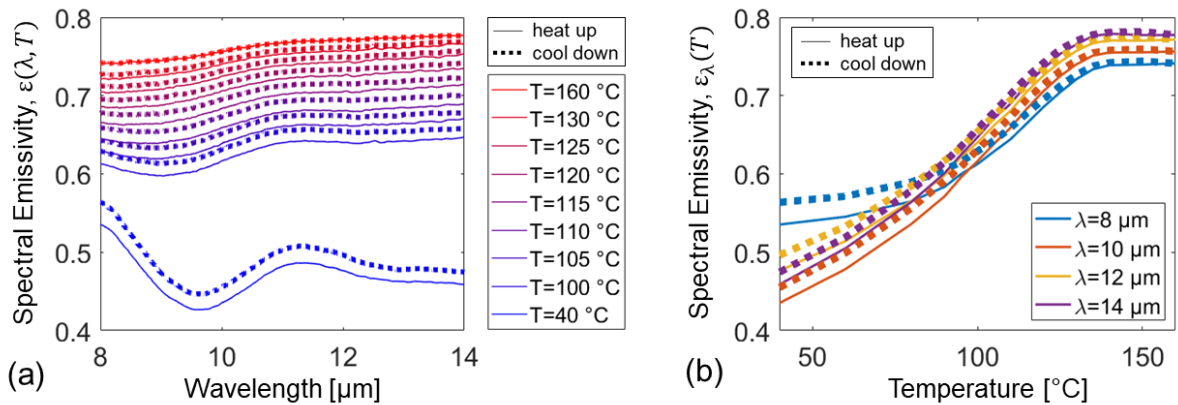

**Figure S4) Emissivity of structure measured during the second thermal cycle. (a)** Spectral emissivity across the spectrum and across transition temperatures. **(b)** Emissivity at discrete wavelengths across transition temperatures

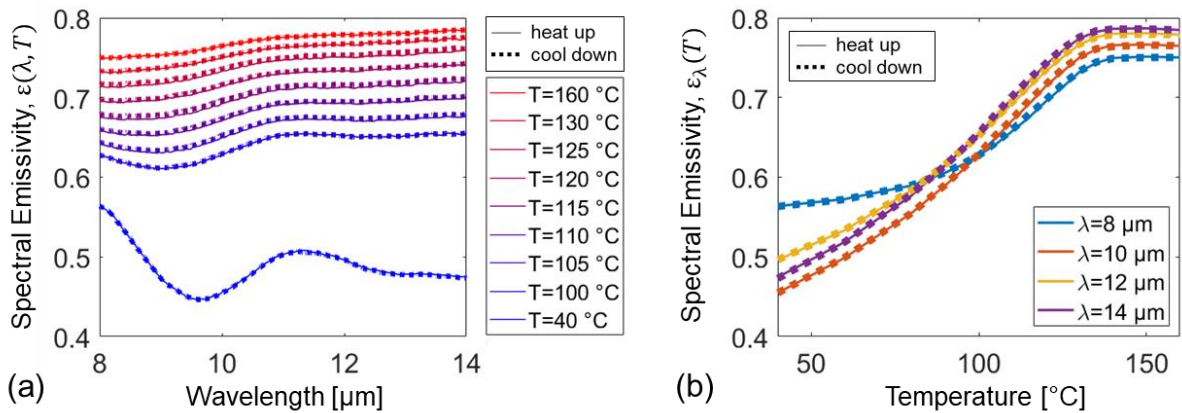

**Figure S5) Emissivity of structure measured during the third thermal cycle. (a)** Spectral emissivity across the spectrum and across transition temperatures. **(b)** Emissivity at discrete wavelengths across transition temperatures

## 7. IR camera measurements of the ZDSE

While measuring our ZDSE, we considered various image zoom levels; i.e., when imaging the sample, we performed analysis on either the entire sample or some subset of the sample. Infrared images of the samples in their entirety are provided in Fig. S6(a) using an apparent temperature scale from 90 to 105 °C for fine detail (instead of 90 to 130 °C scale used in Fig. 3(c) of the main text). Note the large triangular region in the present sample has no SmNiO$_3$ film and is therefore just ITO on glass. We used the largest continuous square coated area for the main text results (marked by red boxes in Fig. S6(a) and (b)) which turned out to be 25×25 pixels, and a second region consisting of a 14×14 pixels square (shown in black in Fig. S6(a) and (c)) that deliberately avoided extraneous speckles within the 25×25 pixels region in order to gauge how well the present sample could perform if free of obvious defects.

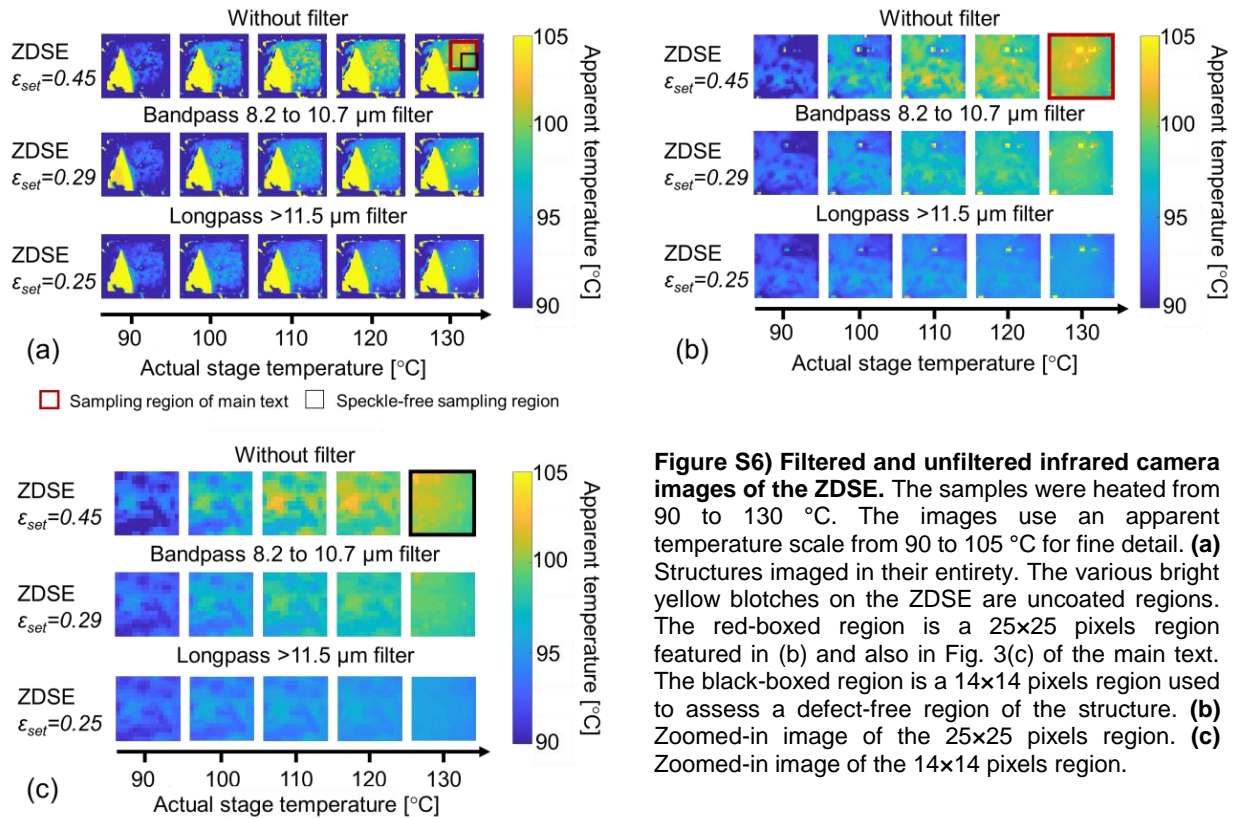

**Figure S6) Filtered and unfiltered infrared camera images of the ZDSE.** The samples were heated from 90 to 130 °C. The images use an apparent temperature scale from 90 to 105 °C for fine detail. **(a)** Structures imaged in their entirety. The various bright yellow blotches on the ZDSE are uncoated regions. The red-boxed region is a 25×25 pixels region featured in (b) and also in Fig. 3(c) of the main text. The black-boxed region is a 14×14 pixels region used to assess a defect-free region of the structure. **(b)** Zoomed-in image of the 25×25 pixels region. **(c)** Zoomed-in image of the 14×14 pixels region.

The pixel-averaged differences in apparent temperature for the 25×25 pixels region and the 14×14 pixels region are given in Table S1.

**Table S1.** Pixel-averaged changes in apparent temperature from 90 °C to 130 °C, across the "main text" sampling region (red square, 25 × 25 pixels) and the "defect-free" sampling region (black square, 14 × 14 pixels)

| ZDSE Region | No Filter | Bandpass 8.2 to 10.7 µm | Longpass >11.5 µm |
|---|---|---|---|
| 25 × 25, red square | 9.6 | 6.9 | 3.0 |
| 14 × 14, black square | 8.3 | 6.3 | 2.5 |